# Deep-learning-based Breast CT for Radiation Dose Reduction

Wenxiang Cong[a], Hongming Shan[a], Xiaohua Zhang[b], Shaohua Liu[b], Ruola Ning[b], Ge Wang[a*]

[a]Biomedical Imaging Center, Department of Biomedical Engineering, Rensselaer Polytechnic Institute, Troy, New York 12180; [b]Koning Corporations, West Henrietta, New York 14586

## ABSTRACT

Cone-beam breast computed tomography (CT) provides true 3D breast images with isotropic resolution and high-contrast information, detecting calcifications as small as a few hundred microns and revealing subtle tissue differences. However, breast is highly sensitive to x-ray radiation. It is critically important for healthcare to reduce radiation dose. Few-view cone-beam CT only uses a fraction of x-ray projection data acquired by standard cone-beam breast CT, enabling significant reduction of the radiation dose. However, insufficient sampling data would cause severe streak artifacts in CT images reconstructed using conventional methods. In this study, we propose a deep-learning-based method to establish a residual neural network model for the image reconstruction, which is applied for few-view breast CT to produce high quality breast CT images. We respectively evaluate the deep-learning-based image reconstruction using one third and one quarter of x-ray projection views of the standard cone-beam breast CT. Based on clinical breast imaging dataset, we perform a supervised learning to train the neural network from few-view CT images to corresponding full-view CT images. Experimental results show that the deep learning-based image reconstruction method allows few-view breast CT to achieve a radiation dose <6 mGy per cone-beam CT scan, which is a threshold set by FDA for mammographic screening.

## 1. INTRODUCTION

X-ray mammography is a routine imaging procedure for breast cancer screening in United States [1]. This technique helps identify the tumor, differentiate benign or malignant, determine the prognosis, and inform the treatment suggestions [2]. It is cost-effective, fast, and of insignificant ionizing radiation dose. However, mammography suffers from the major limitation that a 3D breast structure is projected into a plane [3]. Lesions can be obscured by overlaying tissue structures and cause a false negative. Dense glandular tissue located above and below a lesion can reduce the visibility of the lesion. Digital tomosynthesis produces three-dimensional breast images with high spatial resolution in the transverse plane for breast cancer screening [4]. In tomosynthesis, the x-ray tube moves along an arc over a compressed breast to acquire multiple x-ray projection of the breast over a limited scan angle. The tomosynthetic images reduce the problems caused by tissue overlapping and do allow 3D perception and analysis [5]. The breast tomosynthesis offers improved diagnostic and screening accuracy, and 3D lesion localization. However, digital tomosynthesis collects projections over very limited angles, resulting in spatial resolution loss in the direction perpendicular to the detector [6]. The resolution loss and non-isotropic resolution of digital tomosynthesis limit its ability to detect subtle breast lesions and calcification clusters. In addition, the compression procedure for both mammography and tomosynthesis causes intense persistent pain to the patients and has been associated with other breast conditions including rupture of implants and cystic masses, and hematoma.

Cone beam breast computed tomography (CT) is an emerging x-ray imaging modality that can provide true 3D breast images with isotropic resolution and high-contrast technology, enabling detecting calcifications, discerning and characterizing subtle lesions [7, 8]. The advantages of cone-beam CT include true 3D imaging of the uncompressed breast, reduced motion artifacts, breast-only exposure to radiation, greater efficiency in use of the x-ray beam, no overlapping structures in the breast, and high contrast resolution [9]. CT reconstruction technology offers high-quality images and valuable diagnostic results. However, breast is superficial and sensitive to radiation doses in women. It is important for healthcare to reduce radiation exposure and potential harm to breast [10]. When a low level of radiation dose is distributed over a large number of projection views in breast CT, data acquired at each projection view would have a low signal-to-noise ratio (SNR), which would compromise the quality of image reconstruction. Hence, acquisition of few-view projections in the breast tomographic imaging system can significantly enhance SNR of projection data and reduce radiation dose.

*wangg6@rpi.edu

Image reconstruction of CT from sparse projection data is an ill-posed inverse problem. Direct application of analytic image reconstruction algorithm such as filtered-backprojection (FBP) would lead to poor image quality and severe streak artifacts of image reconstruction. The regularization method of total variation (TV) minimization in compressive sensing framework has been applied for the image reconstruction of few-view CT [11-13]. Such regularization approaches are to maximize data fidelity and image sparsity balanced by regularization parameters. So, the reconstructed image quality heavily depends on regularization parameters. TV regularization method is effective only for reconstruction of piecewise constant images and would over-smoothen textured regions for larger regularization parameter, which may sacrifice important details. The computation cost is still a burden, even though it may be accelerated by parallel computing techniques.

Deep learning is powerful to enhance image quality by reducing image noise and artifacts, and neural network modeling has been extremely successful in image classification, identification, super-resolution imaging, and denoising [14-17]. It may find a regular pattern through learning and inference from a large amount of dataset to reflect real features faithfully and perform various types of intelligence-demanding tasks reliably against uncertainties in system and imaging models [18, 19]. With the composition of multi-layer transformations, very complex non-linear functions can be learned [18]. In this paper, we propose a deep-learning-based image reconstruction approach for the breast computed tomography (CT). Our method directly learns an end-to-end mapping between CT images reconstructed from few-view projections and corresponding CT images reconstructed from full-view projections. The mapping is modeled as a residual neural network (ResNet) through supervised learning based on clinical breast imaging dataset to realize few-view image reconstruction of breast CT. Standard breast CT acquires 300 projection views, and achieves the average mean glandular dose ranges between 7 and 13.9 mGy [20]. This proposed method only requires a fraction of x-ray projection views of standard breast CT and utilizes deep-learning-based method for the image reconstruction, allowing for a significant reduction of radiation dose and achieving the image quality comparable that of standard breast CT.

## 2. DEEP RESIDUAL LEARNING

### 2.1. CNN network

The universal approximation theorem states that neural networks are capable of representing a wide variety of interesting functions through data training [18]. Few-view CT image reconstruction based on learning method is to model a nonlinear mapping from low-dose images $\tilde{I}$ to routine-dose images $I$. The modeling can be implemented by a neural network to establish a non-linear mapping function $m$ denoted by $I = m(\tilde{I}, \theta)$, which parameters $\theta$ are learned from clinical dataset through the neural network model.

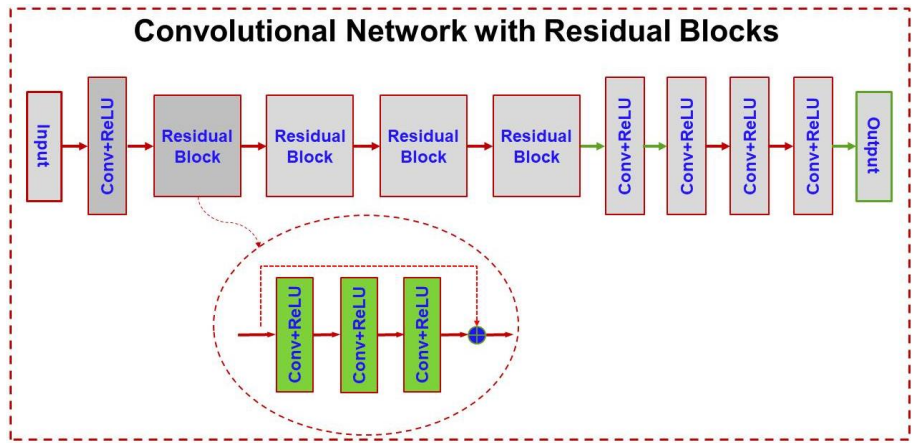

**Fig.1.** Network architecture for few-view breast CT

CNN is a popular structure of neural network for image processing [21]. It is composed of several convolution layer and activation functions. The convolutional layer performs convolution operation on filter kernels and its input data to produce output results as input signals to the next layer. After convolutional operations, activation function is applied to introduce non-linearity. Through data training, CNN network is to find features for a specific task with minimization of the cost function. However, a deep CNN often suffers from vanishing gradient and the gradient divergence problems, which hampers the transmission of information from shallow layers to deep layers. Here, we apply the residual neural network (ResNet) to modeling the non-linear mapping. ResNet is an advanced network architecture, which has the ability to obtain high-accuracy results for image processing [22, 23]. With use of identity shortcut, ResNet alleviates overfitting, reduces vanishing gradient problem, and allows the neural network converge faster and more efficiently. The deep training network enables the extraction of more complex and detailed features from CT image dataset.

## 2.2. Network architecture

Our residual network consists of one convolution layer with 64 filters of 3 × 3 kernels, followed by four residual blocks, two convolution layers with 64 filters of 3 × 3 kernels, and one convolution layer with 32 filters of 3 × 3 kernels. Each layer is followed by a ReLU activation function. The last layer generates only one feature map with a single 3 × 3 filter as the output. Each residual block performs feed forward neural networks with shortcut connections that skip one or more layers. Fig. 1 presents the network architecture. The network is trained using image patches. Training of a network is a process of finding kernels in convolution layers to minimize differences between output predictions and given ground truth labels on a training dataset. A network model performance under particular kernels and weights is calculated by a loss function through forward propagation on a training dataset, and learnable parameters are updated according to the loss value through an optimization algorithm called backpropagation, which uses the chain rule to speed up the computation of the gradient for the gradient descent (GD) algorithm.

## 3. EXPERIMENTS

### 3.1. Two-third radiation dose reduction

Dataset of breast patients are obtained from Koning Corporations, a famous company manufactured breast CT system in America. In clinic, a standard breast CT acquires 300 projection views over 360 degrees for each breast in one scan. Full-view breast CT images are reconstructed from the 300 projection data. In this study, we perform downsampling for 300 projection data of each patient at one third sampling rate to 100 projection views, achieving two third radiation dose reductions for the few-view breast CT. Then, the image reconstruction is performed from full views projection data and 100 projection data using filtered backprojection (FBP) method to obtain full-view and few-view breast CT images, respectively. These data are used to generate ideal supervised training data for accurate neural network modeling.

The standard training cycle is followed through the training, validation and testing stages. Datasets are typically split into training, validation, and testing dataset. A training dataset is used to train a network, where loss function is calculated via forward propagation and learnable parameters are updated via backpropagation. A validation dataset is used to evaluate the model during the training process. A testing dataset is ideally used only once at the very end of the project in order to evaluate the performance of the final model. In our experiments, we extracted overlapping patches of 64 × 64 from 2067 full-view and few-view CT images, obtaining 1,986,387 pairs of image patches as inputs and labels in the training, and 794,554 pairs of patches for validation. During the training phase, the Adam optimization method is used to train the model with a mini-batch of 128 image patches for each iteration. The learning rate is selected to be $1.0\times10^{-4}$. We apply the adjustable learning rate (lR) strategy ( $IR = IR/\sqrt{1+epoch}$ ) for avoiding gradient exploding and speeding up training.

In early epochs, setting relatively high learning rate is beneficial for accelerating training process. The network modeling is implemented in Python with the Tensorflow library, running on a NVIDIA Titan X GPU with 12 GB memory. Data are randomly sampled in the training dataset, maximizing the probability of finding the global minimum of loss function. We calculate the values loss function over the image patches for each epoch to verify the convergence of the optimization. Fig. 2 shows the averaged Manhattan norm loss versus the number of epochs. After network training, we perform the testing for another 942 few-view breast CT images. Our results have shown that the proposed ResNet can effectively reduce noise and artifacts, and preserve structure details of breast images, as shown in Fig.3. The peak signal-to-noise ratio (PSNR) and structural similarity (SSIM) are utilized to quantitatively evaluate the performance of the deep learning image processing methods, achieving PSNR of 34.90 and SSIM of 0.9699 for representative slices.

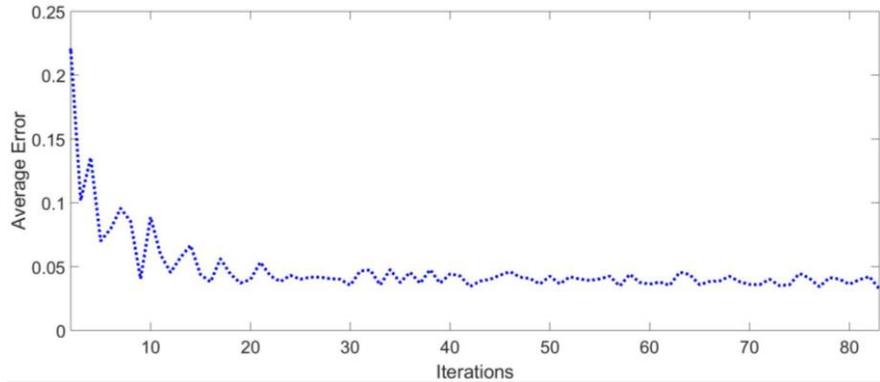

**Fig. 2.** Convergence of neural network described by Manhattan norm loss versus the number of epochs during the training.

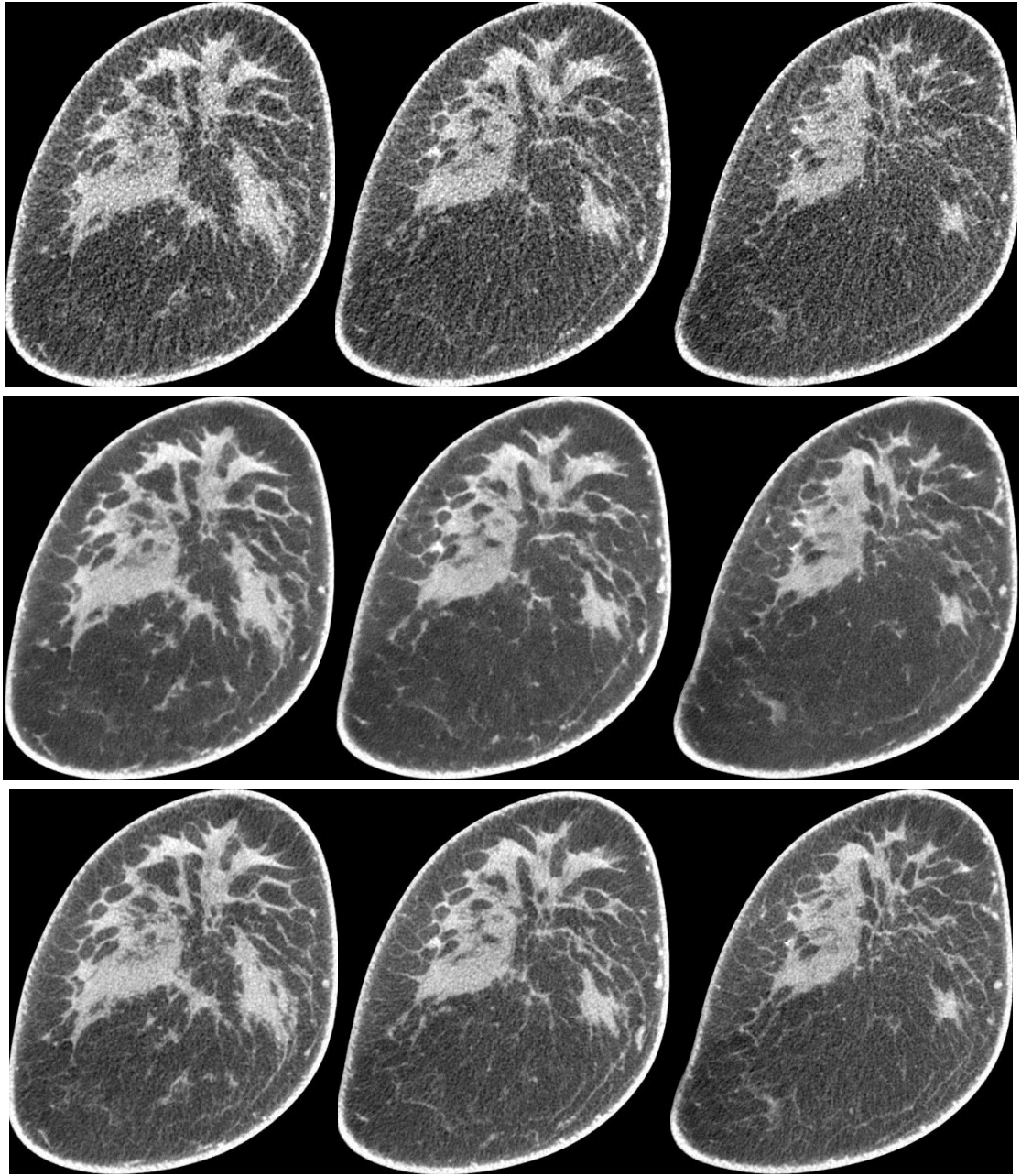

**Fig.3**. Breast CT slices of a patient: 348th slice (first column), 363th slice (second column), and 378th slice (third column) for the evaluation of image reconstruction. CT images in upper row are reconstructed from 100 projection views using FBP method. CT images in middle row are reconstructed from 100 projection views using the deep learning-based image reconstruction method. CT images in lower row are reconstructed from 300 projection views using FBP method.

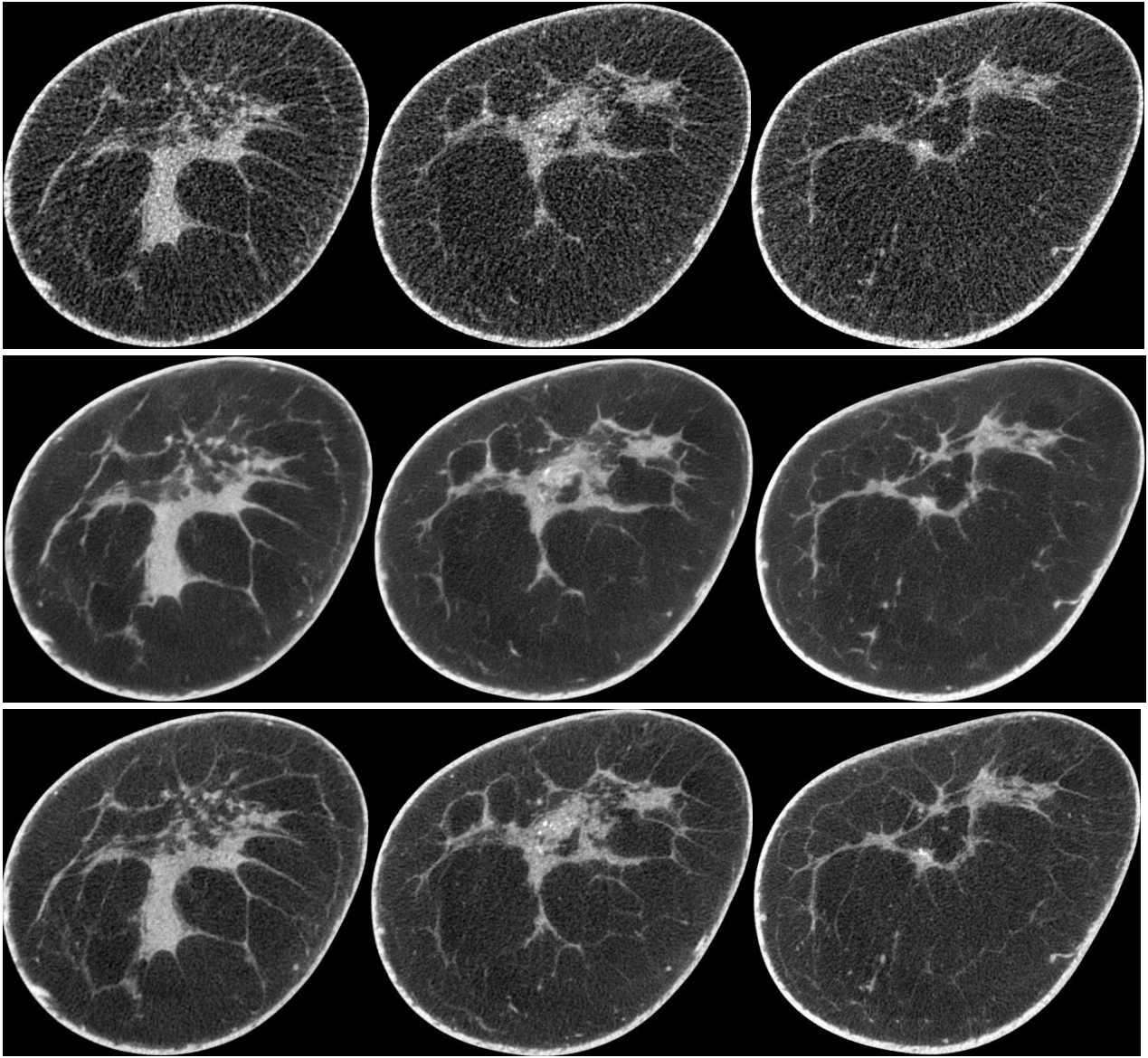

**Fig.4**. Breast CT slices of another patient: 175th slice (first column), 225th slice (second column), and 275th slice (third column) for the evaluation of image reconstruction. CT images in upper row are reconstructed from 75 projection views using FBP method. CT images in middle row are reconstructed from 75 projection views using the deep learning-based image reconstruction method. CT images in lower row are reconstructed from 300 projection views using FBP method.

### 3.2. Three-quarter radiation dose reduction

Furthermore, we perform downsampling for 300 projection data of each patient at a quarter sampling rate to 75 projection views, achieving three quarter radiation dose reductions for the few-view breast CT. Then, the image reconstruction is respectively performed from full views projection data and 75 view projection data using filtered backprojection (FBP) method to obtain full-view and few-view breast CT images. From breast CT images of six patients, we obtain 2,383,664 pairs of image patches as our training inputs and labels for supervised training. From the training dataset, the ResNet neural network is trained to obtain a network model for the reconstruction of few-view CT images. We perform the testing for few-view breast CT images from three patients. Qualitatively, the trained neural network produces high-quality breast CT image from few-view breast CT images, achieving PSNR of 32.96 and SSIM of 0.9494, as shown in Fig.4.

# 4. CONCLUSION

In this study, we developed a ResNet network for the image reconstruction of the few-view breast CT using Python with the Tensorflow library. The ResNet network is trained, validated, and tested based on real clinical breast CT dataset from Koning Inc. Our experimental results show that the optimization model of the neural network is highly cost-effective, and has an excellent convergent behavior in the learning process. The ResNet produces an output of high-quality breast CT images, effectively removing noise and artifacts and preserving structure details of breast images. For the image reconstructions with one third and one fourth of radiation dose delivered by a standard breast CT, we quantitatively evaluate breast images of output from network by the peak-to-noise ratio (PSNR) and structural similarity (SSIM). The proposed network model produces promising results and has an important application value in clinical breast imaging.